# Enumeration Order complexity Equivalency

**Saeed Asaeedi**

(AmirKabir University of Technology, Tehran, Iran

sa_saeed_sa@aut.ac.ir)

**Farzad Didehvar**

(AmirKabir University of Technology, Tehran, Iran

didehvar@aut.ac.ir)

**Abstract:** Throughout this article we develop and change the definitions and the ideas in [1], in order to consider the efficiency of functions and complexity time problems. The central idea here is effective enumeration and listing, and efficiency of function which is defined between two sets proposed in basic definitions.

More in detail, it might be that $h$ and $g$ were **co-order** but the velocity of them be different.

**Introduction**: Every recursive set could be ordered increasingly and every c.e set which could be ordered increasingly is a recursive set , so  increasing order type is related tightly to recursive sets.

 In [1] the posing question is: What about the other c.e sets? Which order types could be associated to an arbitrary c.e set? In that article trying to answer the above question, we define some equivalence relations and we investigate different properties of that. The material in [1] could be considered In Computability field setting.  We could develop the same ideas and approximately in a parallel way in Complexity Theory. So, we define different concepts to cope with some difficulties which arise.

Here, we repeat the definition 1, 2 in [1], with some essential modifications.

**Definition 1:** A listing of an infinite c.e set $A \subseteq \mathbb{N}$ is a bijective computable function $f : \mathbb{N} \to A$.

**Definition 2:**
1. Two listings $h, g$ are **co-order**, $h \sim g$, if $h(i) < h(j) \iff g(i) < g(j)$ for all  $i, j \in \mathbb{N}$.
2. Two c.e subsets A and B of $\mathbb{N}$ with equal cardinality are **co-order**, A~B, if there exist listings $h$ of $A$ and $g$ of $B$ such that $h \sim g$.

**Definition 3:** Let $h$ enumerate $A$. In the case that Turing machine $M$ computes $h$, $Time(h)(n)$ be the essential steps to halt Turing machine $M$. we have:

$$Time(h)(n) = \text{ the number of steps which } M \text{ halts on input } n$$

We apply $Time(h)(n)$ to compare the time of two different listing.

**Definition 4:** We know listing $h$ strictly more rapid than listing $g$ if for any $n \in \mathbb{N}$ we have:

$$Time(h)(n) < Time(g)(n)$$

**Definition 5:** We know listing $h$ more rapid than listing $g$ if there exists $M \in \mathbb{N}$ such that:

$$\forall n > M \quad \sum_{i=1}^{n} Time(h)(i) < \sum_{i=1}^{n} Time(g)(i)$$

**Definition 6:** Let $A$ be a c.e set. $TIME(A)$ represents time complexity of enumerating the set $A$. By supposing $ALL$ as the set of all listing of $A$, It is defined as follows:

$$TIME(A) \in O\big(t(n)\big) \quad iff \; \exists h \in ALL \; such \; that$$

$$h \text{ has an } O\big(t(n)\big) \text{ time deterministic Turing machine}$$

**Definition 7:** Let $A$ be a c.e set. By supposing $ALL$ as the set of all listing of $A$, $NTIME(A)$ is defined as follows:

$$NTIME(A) \in O\big(t(n)\big) \quad iff \; \exists h \in ALL \; such \; that$$

$$h \text{ has an } O\big(t(n)\big) \text{ time non} - deterministic \text{ Turing machine}$$

**Example1:** Let $A$ be the set of all prime numbers, as we know there are infinite numbers of algorithms to produce prime numbers. Since $A = PRIME \in P$, there is a deterministic algorithm for this problem in polynomial time. Consequently, there is a deterministic Turing machine which enumerates $A$ in polynomial time, in other words $TIME(A) \in O(n^k)$.

**Example 2:** Note that $B = SAT$ is a $NP - complete$ problem, so there is a non deterministic Turing machine $M$ which it enumerates $B$ in polynomial time, equivalently $NTIME(B) \in O(n^k)$.

**Remark:** It is notable to know that there are non recursive c.e sets like $A$ such that $TIME(A) \in O(n^k)$, and there are non recursive c.e set like $B$ such that, $NTIME(B) \in O(n^k)$.

**Definition 8:** The c.e set $A$ is **P co-order** if there are sets $B$ and $k \in \mathbb{N}$ such that $TIME(B) \in O(n^k)$ and $A \sim B$.

**Definition 9:** The c.e set $A$ is **NP co-order** if there are sets $B$ and $k \in \mathbb{N}$ such that $NTIME(B) \in O(n^k)$ and $A \sim B$.

**Theorem 1**: Any P co-order set is NP co-order set.

**Proof:** Straight forward.

**Theorem 2:** Any recursive set is P co-order set.

**Two equivalence relations PU and NPU:**

**Definition 10:** $A$ is non deterministic polynomial reducible to $B$ ($A <_{np} B$) if there is computable function $f$, such that there is a Turing machine $M_f$ non deterministic and halts in polynomial time, such that:

$$x \in A \iff f(x) \in B$$

**Definition 11:** Two sets $A$ and $B$ are non deterministic polynomial equivalent ($A \equiv_{np} B$) if $A <_{np} B$ and $B <_{np} A$.

**Definition 12:** Two sets $A, B \subseteq \mathbb{N}$ are **PU** equivalent if $A \sim B$ and $A \equiv_p B$.

**Definition 13:** Two sets $A, B \subseteq \mathbb{N}$ are **NPU** equivalent if $A \sim B$ and $A \equiv_{np} B$.

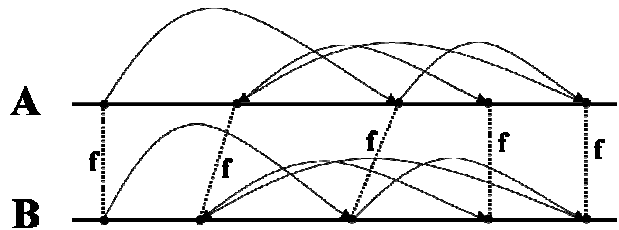

**Conclusion 1:**

A $\equiv_{pu}$ B $\quad \Rightarrow \quad$ A $\equiv_T$ B

**Conclusion2:** Let $A$ and $B$ belong to two different Turing classes. $A$ and $B$ are not **PU** equivalent.

**Conclusion3:** Suppose that for two different subsets of $\mathbb{N}$, $A$ and $B$, A $\in$ P and B $\in$ NP$-$complete. $A \equiv_{pu} B$ iff $P = NP$.

**Theorem 3:** Let $A$ and $B$ be two different subsets of $\mathbb{N}$. $A \equiv_{pu} B$ concludes $A \equiv_{npu} B$

**Lemma 1:** If $P = NP$ and $A \equiv_{np} B$ , we have $A \equiv_p B$.

**Theorem 4:** **NPU** and **PU** are equivalent unless $P \neq NP$.

**Theorem 5:** $P \neq NP$ unless **NPU** and **PU** are equivalent.

*References:*

[1] Safilian, Ali Akbar; Didehvar, Farzad. *Enumeration Order Equivalency.* 2010

[2] Safilian, Ali Akbar; Didehvar, Farzad. "Two new degrees based on enumeration orders and their equivalency relations". 6[th] International Conference of Computability, Complexity, and Randomness (2011) . Cape town, South Africa.

[3] M. Sipser: Introduction to the Theory of Computation. PWS Publishing company. 1997.

[4] S. Cooper. Computability Theory. Chapman & Hall, 2004.